\documentclass[aps,amsmath,amssymb,nofootinbib]{revtex4}
\usepackage{graphicx}
\usepackage{psfrag} 

\def\state#1{| {#1} \rangle}
\def\deriv#1{\frac{\partial}{\partial {#1} }}
\def\half{\frac{1}{2}}
\DeclareMathOperator{\cotg}{cotg}

\begin{document}

\title{Dimensional reduction on a sphere}

\author{Gunnar M\"oller${}^{a}$, Sergey Matveenko${}^{b}$ and  St\'ephane Ouvry${}^a$}
\affiliation{
${}^a$Laboratoire de Physique Th\'eorique et Mod\`eles Statistiques, 91406 Orsay, France\\
${}^b$Landau Institute for Theoretical Physics, Kosygina Str. 2, 119334, Moscow, Russia
}
\date{September 21$^\text{st}$ 2005}

\begin{abstract}

The question of the dimensional reduction of two-dimensional (2d) quantum
models on a sphere to one-dimensional (1d) models on a circle is adressed. A
possible application is to look at a relation between the 2d anyon model and
the 1d Calogero-Sutherland model, which would allow for a  better
understanding of the connection between  2d anyon exchange statistics and
 Haldane exclusion statistics. The latter is realized microscopically in
the 2d LLL anyon model and in the 1d Calogero model. In a harmonic well of
strength $\omega$ 
or on a circle of radius $R$ 
-- both parameters $\omega$ and $R$ have to be viewed as
long distance regulators -- the Calogero spectrum is discrete. It is well
known that by confining the anyon model in a 2d harmonic well and projecting
it on a particular basis of the harmonic well eigenstates, one obtains the
Calogero-Moser model. It is then natural to consider
the anyon model on a sphere of radius $R$ and look for a possible dimensional
reduction to the Calogero-Sutherland model on a circle of radius $R$. First,
the free one-body case is considered, where a mapping from the 2d sphere to
the 1d chiral circle is established by projection on a special class of
spherical harmonics. Second, the $N$-body interacting anyon model is
considered : it happens that the standard anyon model on the
sphere is not adequate for dimensional reduction. One is thus lead to define a
new spherical anyon-like model deduced from the Aharonov-Bohm problem on the
sphere where each flux line pierces the sphere at one point and exits it at
its antipode.
\end{abstract}

\maketitle

\section{Introduction}
\label{sec:intro}

There has been some interest in relating the two dimensional anyon
model \cite{Leinaas77,Goldin81,Wilczek82a,Wilczek82b} to the one dimensional
Calogero model \cite{Calogero69,Calogero71,Sutherland71,Sutherland72}. These
studies \cite{Hansson92,Brink93} were motivated by the fact that both models
describe particles with non standard statistics, which
interpolate from Bose-Einstein statistics to Fermi-Dirac statistics. In the
anyon case, exchange statistics is at work, based on the non trivial braiding
properties of the $N$-body configuration space; in the Calogero case,
Haldane-exclusion statistics is manifest \cite{Haldane91}, based on Hilbert
space counting arguments. Although a priori different, a non ambiguous
relation has been established \cite{OuvryMapping,OuvryMacris,Ouvry} between the anyon
model and the Calogero model using as long distance regularisation a
confining harmonic well: by projecting the anyon model on a
particular subspace of the 2d harmonic well Hilbert space, one obtains the 1d
Calogero model in a harmonic well (Calogero-Moser model). This mapping relates
Haldane-exclusion statistics to anyon-braiding statistics, as
already foreseen in the LLL-anyon model \cite{OuvryLLL,OuvryLLLbis}, where
Haldane thermodynamics are realized microscopically. A more intuitive way to
look at the harmonic projection is to notice that the 2d harmonic quantum
numbers on which the projection is made would be those of the lowest Landau
level if a magnetic field were present. In other words, a zero magnetic field
limit has been taken, which indeed becomes meaningful via the harmonic well
regularisation : it follows that the vanishing magnetic field limit of the
LLL-anyon model is the Calogero model \cite{OuvryMapping}.

The present work discusses whether a similar correspondence can be established
with a different regularisation scheme, for example by modifying
the topology of the 2d plane to a sphere (see also a previous attempt to
dimensionally reduce Laughlin wave functions on a cylinder \cite{Haldane-Reza}).
One should obtain a
compact anyon model with a discrete spectrum, which might be dimensionally
reducible to the Calogero-Sutherland (C-S) model on the circle. Note  that 
a direct relation between the Calogero-Moser
and the C-S models has already been discussed in Ref.\  \onlinecite{Nekrasov97}.
The C-S model
is a prime example of a solvable model with exclusion statistics, and
therefore its relation to the anyon model seems natural. This motivates our
respective analysis of the anyon model on a sphere of radius $R$, which, by projection on a special
class of spherical harmonics, might yield the C-S model on a
circle of the same radius.

As a first step, the free cases of a quantum mechanical particle on the circle
and on the sphere are analyzed in section \ref{sec:non_interacting}. A
dimensional reduction scheme from the sphere to a chiral circle is achieved,
which mimicks the harmonic dimensional reduction scheme. As a second step, the
question of defining the anyon model on a sphere is adressed. The anyon model
on the sphere has been discussed by various
authors \cite{Lee89,Lechner92,OuvryComtetMcCabe,Li93,Park94}, yet basic
differences with regard to the C-S model immediately show up.
In the latter model, the statistical parameter is a continuous coupling
parameter, while in the spherical anyon model, only discrete
$N$-dependent statistical parameters are allowed due to Dirac quantization of
the total flux on the sphere. This restriction may also be regarded as a
property of the braid group on the sphere, since a loop made by a particle
encircling all others is contractible on the sphere and has trivial braiding
properties. Another important difference lies in the scaling of the spectrum
with the statistical parameter. In the anyon model, the integrable part of the
spectrum scales linearly \cite{OuvryComtetMcCabe}, whereas the spectrum of the
C-S model is quadratic. One is thus lead to propose a new
anyon-like model on the sphere in section \ref{sec:anyon_proposal}, whose
properties will be discussed. Finally, section \ref{sec:conclusion} gives a
summary of the results obtained so far.

\section{Non interacting cases}
\label{sec:non_interacting}
\subsection{A reminder: harmonic dimensional reduction}
\label{sec:A_reminder}
In the free case, it is known how to map a 2d particle in a harmonic well on a
1d particle in a harmonic well. One starts from the 2d Hamiltonian
 \begin{equation}
\label{ham} H=-2{\deriv{\bar z}} {\deriv{z}}+\frac{1}{2}\omega^2z\bar z
 \end{equation}
with spectrum
\begin{equation} E_{nm}=\omega(2n+|m|+1),\end{equation}
where $n\in \mathbb{N}$ is the radial quantum number and $m\in \mathbb{Z}$ is
the orbital quantum number. Specializing to the particular subset of quantum
numbers, i.e., on each degenerate energy level one picks up the state of
maximal angular momentum $l\ge 0$ and radial quantum number $n=0$
 (notice that these are the LLL
quantum numbers if a magnetic field were present)
\begin{equation}  \label{88}
\langle z,\bar z|0,l \rangle =z^{l}
e^{-\half\omega z\bar z},\quad
l\ge 0\end{equation}
with spectrum
\begin{equation}  E_{l}=\omega(l+1)\end{equation}
and projecting the Hamiltonian on this subspace -- i.e., assuming
$\Psi=f(z)e^{-\half\omega z\bar z}$ -- gives
\begin{equation}\label{ninn} \omega\left(1+  z{\deriv{z}}\right) f(z)= Ef(z).
\end{equation}
Note that one might have as well used the subspace
$n=0$ and $l\le 0$, that is the eigenfunction
$\bar \Psi=f(\bar z)e^{-\half\omega z\bar z}$ to obtain
the eigenvalue equation
\begin{equation}
\label{ninnbar} \omega\left(1+  \bar z\bar\partial\right)f(\bar z)= Ef(\bar z).
\end{equation}
In the thermodynamic limit $\omega\to 0, l\to\infty$ with $\omega l$ fixed,
the physical picture is that of states with a probability density significant
only increasingly close to the edge of the 2d plane, thus mimicking particles
on the 1d boundary of the 2d sample.

(\ref{ninn}) (and (\ref{ninnbar})) is nothing but an eigenvalue equation for a
1d harmonic well Hamiltonian in the coherent state representation. Consider
the 1d harmonic oscillator Hamiltonian
\begin{equation}
  \label{eq:ham_HO}
  H=-\half\left(\frac{d}{dx}\right)^2+\frac{1}{2}\omega^2 x^2 =
\omega\left(a^\dagger a+\frac{1}{2}\right).
\end{equation}
Coherent states $\state{\alpha}$ are eigenstates
of the annihilation operator
\begin{equation}
  \label{eq:ann_HO}
  a=\sqrt{\frac{\omega}{2}}x+\frac{i}{\sqrt{2\omega}}p_x.
\end{equation}
Applying  $a^\dagger$ to the coherent
states $|\alpha\rangle$ in the canonical basis $|n\rangle$
\begin{equation}
  \label{eq:coherent_rep}
  |\alpha\rangle = e^{-\frac{|\alpha|^2}{2}}\sum_{n=0}^\infty \frac{\alpha^n}{\sqrt{n!}}|n\rangle,
\end{equation}
one obtains
\begin{equation}
  \label{eq:act_adagger}
  a^\dagger|\alpha\rangle = \bigl( \deriv{\alpha} + \frac{\bar \alpha}{2} \bigr)|\alpha\rangle.
\end{equation}
This  leads to the Hamiltonian in the  coherent states basis
\begin{equation}
  \label{eq:Ham_OH_coherent}
  H=\omega\bigl(\alpha\deriv{\alpha} + \frac{\alpha\bar\alpha}{2}+\frac{1}{2}\bigr),
\end{equation}
which may be rewritten as  (\ref{ninn}) via the non-unitary transformation
(up to a zero-point energy shift)
\begin{equation}
  \label{eq:tilde_H_OM}
  \tilde H= e^{\frac{|\alpha|}{2}^{\!2}} H\, e^{-\frac{|\alpha|}{2}^{\!2}} =
  \omega\bigl( \half+ \alpha\deriv{\alpha} \bigr).
\end{equation}
The mapping between the canonical states
basis  and the coherent states basis (Bargman transform)  maps
 $\alpha^n$ (eigenstates
in the coherent basis) on the Hermite polynomials $H_n(x)$ (eigenstates in the
configuration  space)
\begin{equation}
  \label{eq:Bargman_transform}
  \alpha^n = \frac{1}{\sqrt{2^n}} \int_{-\infty}^\infty dx\, H_n(x)
   e^{-x^2+\sqrt{2} \alpha x - \frac{\alpha^2}{2} }.
\end{equation}
In the sequel, a similar dimensional reduction is outlined which establishes a mapping
from a 2d particle on a sphere to a 1d chiral particle on a circle.

\subsection{Free particle on the circle}
\label{sec:free_circle}

The Hamiltonian of a
 free particle on a circle is
\begin{equation}
  H=\frac{1}{2mR^2}{\hat L^2},\quad \vec r=\cos\phi \vec e_x + \sin\phi\vec e_y
  \label{eq:H_0_circle}
\end{equation}
where $\hat L=-i\frac{\partial}{\partial \phi}$
is the angular momentum operator
in the coordinate representation. For simplicity of
notations, one sets $m=R=1$. The eigenstates $\state{l}$ with
momentum $l\in Z$ are 
\begin{equation}
  \label{eq:solution_circle_1}
  \langle \phi | l \rangle = \frac{1}{\sqrt{2\pi}} e^{il\phi}
\end{equation}
with
eigenvalues $E_l=\frac{1}{2}l^2$ ($l\in \mathbb{Z}$ enforces canonical
single valued wave functions). One now introduces coherent states on the
circle \cite{Kowalski}, starting
from the unitary operator
\begin{equation}
  \label{eq:unitary_ladder}
  U=e^{i\hat\phi},
\end{equation}
which defines a ladder operator on the  basis $|l\rangle$
\begin{equation}
  \label{eq:ladder_action}
  U|l\rangle = |l+1\rangle,
\end{equation}
since $[\hat L,U]=U$.
By analogy with the coherent
states of the harmonic oscillator, one introduces the operator
\begin{equation}
  \label{eq:coherent_operator}
  \exp\bigl\{i(\hat\phi+i\hat L)\bigr\}=U \exp\Bigl\{-\hat L - \frac{1}{2}\Bigr\}
\end{equation}
and its eigenstates $|\xi\rangle$, which are the  coherent states.
In the canonical basis $\{|l\rangle\}$
\begin{equation}
  \label{eq:coherent_j_representation}
  |\xi\rangle=\sum_l \xi^{-l} e^{-\frac{1}{2}l^2}|l\rangle.
\end{equation}
One then  calculates the action of the angular momentum operator on
 $|\xi\rangle$
\begin{equation}
  \label{eq:j_acts_on_xi}
  \hat L|\xi\rangle = \sum_l l \xi^{-l} e^{-\frac{1}{2}l^2}|l\rangle
   = -\xi\frac{\partial}{\partial\xi}\sum_l \xi^{-l} e^{-\frac{1}{2}l^2}|l\rangle
   = -\xi\frac{\partial}{\partial\xi}|\xi\rangle.
\end{equation}
Thus, the Hamiltonian in the coherent states representation is
\begin{equation}
  \label{eq:H0_coherent} H= \frac{1}{2}
  \left(\xi\frac{\partial}{\partial\xi}\right)^2, \end{equation} 
with eigenstates $\psi_l(\xi)=\xi^l$ and eigenvalues $E_l=\half l^2$ with $l\in Z$.

This mapping from the basis of angular momentum to coherent eigenstates establishes an 
equivalent on the circle of the Bargman transform (\ref{eq:Bargman_transform}) for the 
1d harmonic well. Directly evaluating $\langle l | \xi \rangle$ or inserting the 
completeness relation $I=\frac{1}{\pi}\int d\phi|\phi\rangle\langle\phi|$ gives its analogue 
\begin{equation}
  \label{eq:bargman_equivalent}
  \xi^l = \frac{1}{2\pi^2}\,e^{\half l^2}\int_0^{2\pi} d\phi\, e^{-il\phi}\,
\theta_3\Big(\half(\phi+i\ln\xi)\Big|\Big.\frac{i}{2\pi}\Big).
\end{equation}

\subsection{Free particle on the sphere}
\label{sec:free_sphere}
The Hamiltonian of a free particle on the sphere is again
(\ref{eq:H_0_circle}) but for the 2d  angular momentum  operator of
a particle  confined on a sphere of radius $R$. As above, one sets
$m=R=1$. The eigenstates are  the spherical harmonics
$Y_l^m(\theta,\phi)=\langle \theta,\phi|l,m\rangle$, with $l = 0,1,2,\ldots$
and $-l \leq m \leq l$. Their energies are degenerate in $m$
with the eigenvalue equation
\begin{equation}
  \label{eq:eigenvalues_sphere_0}
  H|l,m\rangle=\frac{1}{2}l(l+1)|l,m\rangle.
\end{equation}
On the circle, the spectrum has a twofold degeneracy on each energy level
(except at $l=0$). One wishes to select a subspace of the 2d Hilbert space
$\bigl\{|l,m\rangle,\, l\in\mathbb{N}, |m|\leq l\bigr\}$ to
establish a mapping on the circle. This subspace has to
be characterized by a single quantum number. It is natural to consider either
$\bigl\{ \bigl|\,l,l\bigr\rangle\bigr\}$
or $\bigl\{ \bigl|\,l,-l\bigr\rangle\bigr\}$
in analogy with the two subspaces discussed for the
harmonic well in \ref{sec:A_reminder}.
In configuration space, these eigenstates rewrite

\begin{equation}
  \label{eq:Yll}
 \langle\theta,\phi|l,l\rangle= Y_l^l(\theta,\phi)=\sin^l\theta\, e^{il\phi}\quad \text{or }
 \langle\theta,\phi|l,-l\rangle= Y_{l}^{-l}(\theta,\phi)=\sin^{l}\theta\, e^{-il\phi}.
\end{equation}
The physical picture is that of states with a  probability density
significant only increasingly close to the equator with increasing angular momentum $l$,
thus mimicking a particle on the equatorial circle.

To achieve the relation with the coherent state representation on the circle, one rewrites
the Hamiltonian on the sphere in stereographic coordinates $(z,\bar z)$, with
$z=\cot(\theta/2)e^{i\phi}$
\begin{equation}
  \label{eq:H_0_stereographic}
  H=-\frac{1}{2}(1+z\bar z)^2\frac{\partial}{\partial z}\frac{\partial}
{\partial \bar z}.
 \end{equation}
The wave functions (\ref{eq:Yll}) rewrite as
\begin{equation}
  \label{eq:Yll_stereographic}
  \langle z,\bar z|l,+l\rangle = \left(\frac{z}{1+z\bar z}\right)^l\equiv u^l\text{, and }\;
  \langle z,\bar z|l,-l\rangle = \left(\frac{\bar z}{1+z\bar z} \right)^l\equiv \bar u^l.
\end{equation}
Note that in the thermodynamic limit, $R\rightarrow\infty$, $u^l$ or $\bar u^l$ reproduces
the lowest Landau states in the presence of a magnetic monople, since if one
reintroduces the
$R$ dependence,
$$
u=\frac{z}{1+\frac{z\bar z}{4R^2}}\stackrel{R\rightarrow\infty}{\longrightarrow}z
\text{, and }\;\bar u\to \bar z.
$$

Projecting on these particular Hilbert spaces $\{|l,+l\rangle\}$ or $\{|l,-l\rangle\}$  means that the Hamiltonian
 acts on functions of $u$ only or $\bar u$ only. One
obtains either

\begin{equation}
  \label{eq:H_in_u}
  H=\frac{1}{2}\left(u^2\bigl(\deriv{u}\bigr)^2+2u\deriv{u}\right)
\end{equation}
or
\begin{equation}
  \label{eq:H_in_v}
  H=\frac{1}{2}\left(\bar u^2\bigl(\deriv{ \bar u}\bigr)^2+2\bar u
\deriv{\bar u}\right).
\end{equation}
These Hamiltonians are basically  (\ref{eq:H0_coherent}), up
to a zero point shift for the energy and the angular momentum
\begin{equation}
  \label{eq:H_shift_uv}
  H_w =\frac{1}{2}\left[\bigl(w\deriv{w}+\frac{1}{2}\bigr)^2-\frac{1}{4} \right]=
  \frac{1}{2}\left[\bigl(\hat L_w+\frac{1}{2}\bigr)^2-\frac{1}{4}\right],
\end{equation}
where $w$ is either $u$ or $\bar u$, and spectrum 
$E_l=\half\big((l+\half)^2-\frac{1}{4}\big)$ with $l\in N$.
Therefore, the chiral Hilbert subspaces allow for a dimensional
reduction from the two dimensional problem on the sphere to the one dimensional
chiral problem on the circle, as can be seen from
(\ref{eq:H_0_circle}) and (\ref{eq:H_shift_uv}), 
with spectra which are both quadratic in the single quantum number of either problem.

\section{A generalized Aharonov-Bohm model on the sphere}
\label{sec:anyon_proposal}
As exposed in the introduction, we are interested in a model on the
sphere, which might be dimensionally reducible to the Calogero-Sutherland 
model. Since the latter features exclusion statistics, a connection with the
anyon model appears possible. Though, the spectrum of the anyon
model as defined in \cite{Leinaas77,Goldin81,Wilczek82a,Wilczek82b}
scales linearly with the interaction parameter, whereas the C-S spectrum
scales quadratically. 

On the plane, anyons are defined either by the nontrivial monodromy
of the  $N$-body wave function with a free Hamiltonian,
or equivalently,  via a singular gauge transformation, by
a monovalued $N$-body wave function (bosonic by convention), with an interacting Aharonov-Bohm
$N$-body Hamiltonian describing a situation where each particle
carries a (statistical) flux line of strength $\Phi=\alpha\Phi_0$
(where $\Phi_0$ is the flux quantum). In this section, we construct a model
on the sphere starting from the Aharonov-Bohm problem and the principle of flux attachment, 
which can be traced back to a geometrical definition of the statistical phase between two particles.

The standard Aharonov-Bohm problem consists of  a
particle coupled to a single flux-tube piercing the plane at the origin. The
Aharonov-Bohm problem on the sphere, considered in Ref.\ \onlinecite{Kretzschmar65},
consists of a particle coupled to a  flux piercing the
sphere at the southpole and exiting  at the northpole.
A spectrum quadratic in $\alpha$ has been obtained,
with a continuous Aharonov-Bohm coupling parameter $\alpha$,
since there is no Dirac quantization
condition, given that the total flux through the sphere is null.

\subsection{Single particle case:  interaction with a flux line $\Phi$
entering the sphere at the South pole and exiting at the North pole}
\label{sec:AB}

Consider a charged particle on a sphere (spherical coordinates : polar angle
$\theta$, azimuthal angle $\phi$). The circulation of the potential vector
along each azimuthal circle (constant $\theta$) has to be equal to the flux
$\Phi$. Therefore
\begin{equation}
\vec A=\frac{\Phi}{2\pi}{\vec{\partial}}\phi.
\end{equation}
The Aharonov-Bohm Hamiltonian $H^{\mathrm AB}$ is obtained by substitution
of the multivalued phase $\exp(i\alpha\phi)$ in the free Hamiltonian
$H^o=-\frac{1}{2R^2}\Delta$ (in  units $\bar h=m=1$), i.e.,
$\Psi_o=\exp(i\alpha\phi)\Psi$, with $\Phi=\alpha \Phi_o$
(up to $\delta$ contact terms at the South and
North poles)
\begin{equation}
  \label{eq:H_AB1}
H^{\mathrm AB}=-\frac{1}{2R^2}\left[ \frac{1}{\sin\theta}\frac{\partial}{\partial\theta}\sin{\theta}
\frac{\partial}{\partial\theta}+\frac{1}{\sin^2\theta}\left(\frac{\partial}{\partial\phi}+i\alpha\right)^2\right].
\end{equation}
The eigenfunctions of (\ref{eq:H_AB1}) are
found to be  \cite{Kretzschmar65} ($x=\cos\theta$)
\begin{equation}
  \label{eq:psi_AB1}
  \Psi = e^{im\phi}P_\lambda^{-\mu}(x),
\end{equation}
where the Legendre polynomials $P_\lambda^{-\mu}(x)$ are related to the hypergeometric function
\begin{equation}
  \label{eq:solution_AB1}
  P_\lambda^{-\mu}(x)=\frac{1}{\Gamma(1+\mu)}\left(\frac{1-x}{1+x}\right)^{\frac{\mu}{2}}
F_H\left(-\lambda, \lambda+1; \mu+1;\frac{1-x}{2}\right)
\end{equation}
and $\mu=|m+\alpha|$, with $m$ the angular quantum number of $L_z$. Eq.\
(\ref{eq:solution_AB1}) yields vanishing boundary conditions for the
wave function, imposed at the South and North poles where the flux line
pierces the sphere, for $\lambda=\mu+k$ with $k$ a positive integer ; the
spectrum is $E_\lambda = \lambda(\lambda+1)$, thus it contains terms quadratic
in $\alpha$, which are also present in the Calogero-Sutherland spectrum.

Stereographically projected coordinates $(z,\bar z)$ on the projective plane
containing the South pole turn out to be useful ($r$ is distance from the
South pole in the projective plane, $\rho=r/(2R)=\cotg(\theta/2)$ is the
rescaled radial coordinate, we set $R=1$):

\begin{equation}
  \label{eq:stereographic_z}
  z=\rho \,e^{i\phi}.
\end{equation}
One has
\begin{align}
  \label{eq:relation_z_theta}
  \cos\theta&= \frac{\rho^2-1}{\rho^2+1},\quad \sin\theta=\frac{2\rho}{\rho^2+1},\quad \rho^2=\frac{1+x}{1-x},\\
\intertext{as well as}
\frac{1-x}{2}&=\frac{1}{ 1+z\bar z}\quad\text{and}\quad \frac{1+x}{2}
  =\frac{z\bar z}{1+z\bar z}.
\end{align}
Looking  at (\ref{eq:psi_AB1}), and starting rather from the free
Hamiltonian (in projective coordinates)
\begin{equation}
\label{}
H^o=-\frac{1}{2}(1+z\bar z)^2\partial\bar\partial,
\end{equation}
the eigenstates are
\begin{equation}
  \label{eq:psi0_NS}
  \Psi^o=e^{i(m+\alpha)\phi}\tan(\theta/2)^{|m+\alpha|}
\frac{1}{\Gamma(1+\mu)}
F_H\left(-\lambda, \lambda+1; \mu+1;\frac{1-x}{2}\right),
\end{equation}
which is either, if $m+\alpha\ge 0$,
\begin{equation}
  \label{ouf}
  \Psi^o=F^{m+\alpha}\frac{1}{\Gamma(1+\mu)}
 F_H\left(-\lambda, \lambda+1; \mu+1;\frac{1-x}{2}\right),
\end{equation}
or if $m+\alpha\le 0$
\begin{equation}
  \label{oufbis}
  \Psi^o={\bar F}^{-m-\alpha} \frac{1}{\Gamma(1+\mu)} F_H\left(-\lambda, \lambda+1; \mu+1;\frac{1-x}{2}\right).
\end{equation}
In (\ref{ouf},\ref{oufbis}), $F=1/\bar z$  simultaneously
encodes on the sphere both the A-B phase and the analogous of a
``short distance'' behavior :
it defines the South-North pole Aharonov-Bohm problem (on the plane,
one has $F=z$ for the A-B
problem with a vortex at the origin).

The appearance of
$F$ in the eigenstates is not accidental.
$\phi$ being an harmonic function on the sphere, the
Cauchy-Riemann equations yield  the function
$F=|F|\exp(i\phi)$:
in the local coordinate system
spanned by the vectors $\partial_\theta\leftrightarrow \partial_x$ and
$\partial_\phi/\sin\theta\leftrightarrow \partial_y$:
\begin{equation}
\label{eq:Cauchy-Riemann}
\partial_\theta \ln|F|=\frac{1}{\sin\theta}\quad\quad\quad
\frac{1}{\sin\theta}\partial_\phi \ln|F|=0
\end{equation}
Therefore, $|F|= \tan\frac{\theta}{2}$ and then $F=\tan\frac{\theta}{2}\exp(i\phi)=1/\bar z$.

As an illustration of (\ref{eq:psi0_NS}),
consider the simple case $k=0$, i.e., $\lambda=|m+\alpha|$, where
the hypergeometric function  rewrites
\begin{equation}
  \label{eq:simple_hypergeom}
  F_H\left(-\lambda, \lambda+1; \lambda+1,\frac{1-x}{2}\right) = \left(1 - \frac{1-x}{2}\right)^\lambda = \left( \frac{1+x}{2}\right)^\lambda=
\left( \frac{z\bar z}{1+z\bar z}\right)^{\lambda}.
\end{equation}
If $m+\alpha\ge 0$
\begin{equation}
\label{u}
  \Psi^o=\left(\frac{z}{1+z\bar z}\right)^{m+\alpha},
\end{equation}
whereas if $m+\alpha\le 0$
\begin{equation}
\label{baru}
  \Psi^o=\left(\frac{\bar z}{ 1+z\bar z}\right)^{-m-\alpha},
\end{equation}
with eigenvalue $E=|m+\alpha|(|m+\alpha|+1)$.
For  $-1/2<\alpha\le 1/2$, the ground state is
either (\ref{baru}) with $m=0$, if $-\frac{1}{2}<\alpha<0$
or (\ref{u}) with $m=0$, if $0<\alpha<\frac{1}{2}$.

One explicitly sees that although $F$ is singular at the South pole, the wave
function is still regular both at the South and North poles -- as it should --
because of appropriate terms in the hypergeometric function.

Proceeding as on the plane, one can bypass the A-B Hamiltonian and define
a new Hamiltonian $\tilde H$, directly obtained from $H^o$ by extracting
$F^{\alpha}$ (or $\bar F^{-\alpha}$) in $\Psi^o$, i.e.,
$\Psi^o=F^{\alpha}\tilde \Psi$, and consequently 

\begin{equation}
\label{Htilde1}
 \tilde H=-\frac{1}{2}(1+z\bar z)^2\bigl( \partial\bar\partial 
                                          - \alpha\frac{1}{\bar z}\,\partial\bigr).
\end{equation}
Or, rather taking advantage of (\ref{u}) (or (\ref{baru})), one can define a
Hamiltonian $\tilde H'$ obtained from $H^o$ by extracting $(\frac{z}{1+z\bar
z})^{\alpha}$ (or $ (\frac{\bar z}{1+z\bar z})^{-\alpha}$), i.e.,
$\Psi^o=(\frac{ z}{ 1+z\bar z})^{\alpha}\tilde \Psi'$, which yields

\begin{equation}
\label{Htildeprime1}
 \tilde H'= -\frac{1}{2}(1+z\bar z)^2\left(\partial\bar\partial + \alpha
\frac{1}{z(1+z\bar z)} \bar \partial - \alpha \frac{z}{1+z \bar z}
\partial\right) +\frac{1}{2} \alpha(\alpha+1).
\end{equation}
Both $\tilde H$ and $\tilde H'$ have simple form.

\subsection{Single particle case : interaction with a flux line $\Phi$ entering the sphere
 at a  point  and exiting at the antipode}
\label{sec:inclined_vortex}

In order to generalize to the $N$-body problem, consider now a
particle at position $z_i$  coupled to a vortex $\Phi$ entering the sphere
at  $z_j$ and exiting at its antipode $-1/\bar z_j$. This problem is
identical to the precedent, the spectrum is the same; it remains to
rewrite the wave functions and the corresponding function $F_{ij}$ which
generalizes $F$ in the appropriate  coordinates $z_i$ and $z_j$.

\begin{figure}[ttt]
\psfrag{aphi0}[c][c]{$\alpha\phi_0$}
\psfrag{theta}[l][c]{$\theta_{ij}$}
\psfrag{phi}[l][c]{$\phi_{ij}$}
\includegraphics[width=0.45\columnwidth]{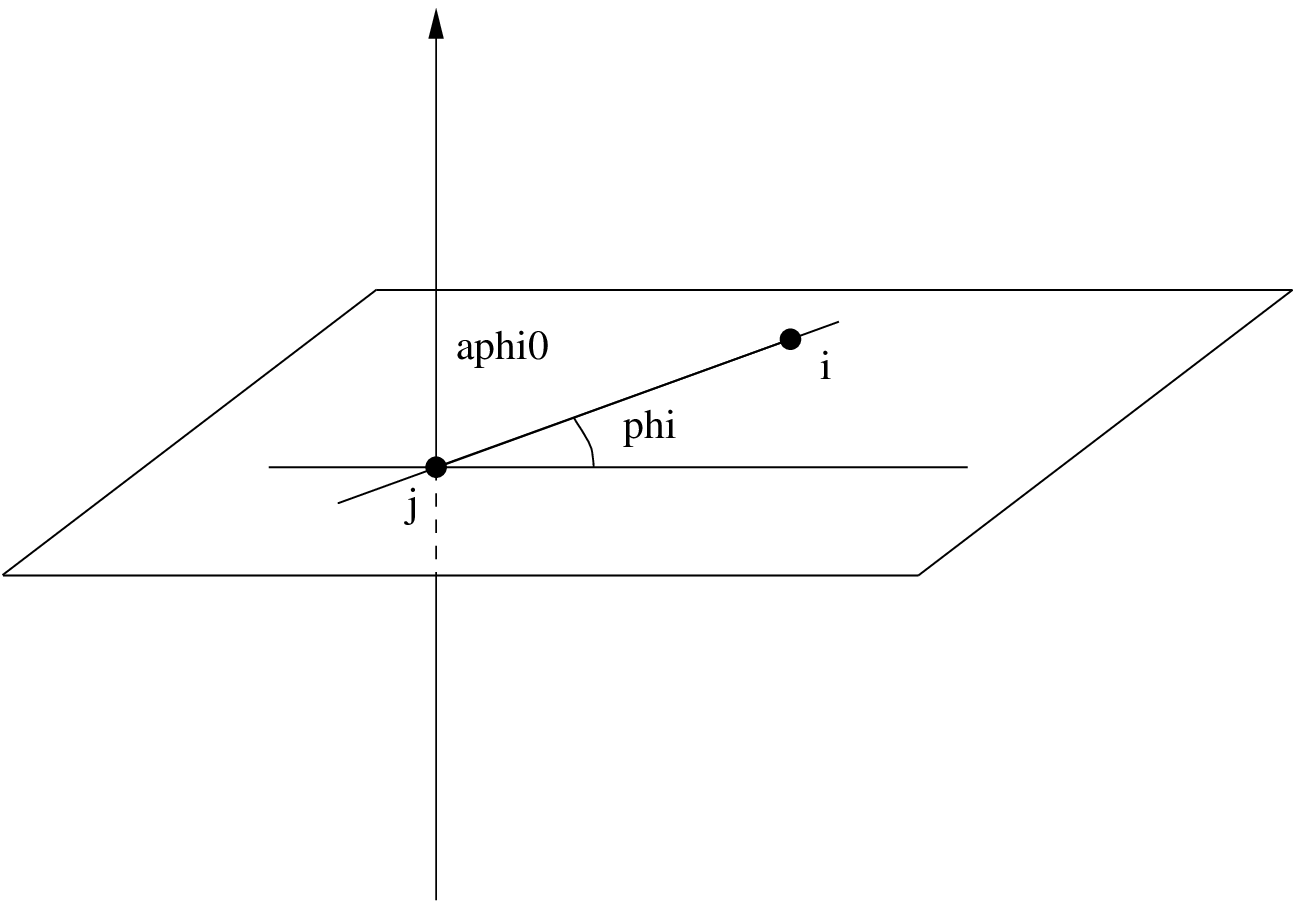}
\hspace{1cm}
\includegraphics[width=0.3\columnwidth]{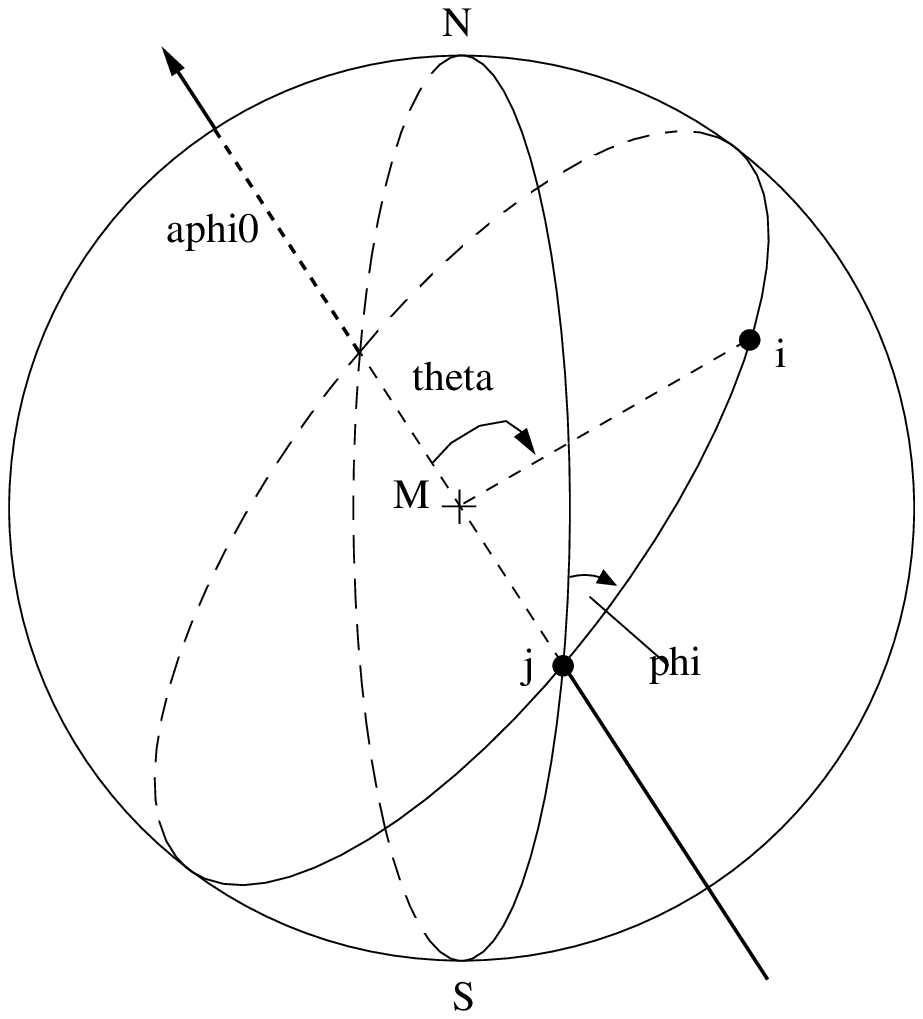}
\caption{
\label{fig:angles}
On the plane, the relative angle $\phi_{ij}$ of a vortex $z_j$ and a particle
  $z_i$ is the angle between a reference geodesic -- i.e., a straight line, here the
  horizontal line -- passing through the vortex $z_j$ and the geodesic passing
  through both the vortex $z_j$ and the particle $z_i$. On the sphere,
  $\phi_{ij}$ is the angle between a reference geodesic -- i.e., a great circle
  of the sphere, here passing through both poles and the vortex $z_j$ --
  and the geodesic passing through both the vortex $z_j$ and
  the particle $z_i$ (it also passes through the antipodes $-1/\bar z_j$,
  $-1/\bar z_i$). Equivalently, one can consider the azimutal angle of the
  particule $i$ in the coordinate system where the South-North axis coincides
  with the flux line $j$ : this is again $\phi_{ij}$, up to a constant.}

\end{figure}

It turns out that the A-B phase    on the sphere has an analogous geometric
interpretation to the one  on the plane as depicted in FIG. \ref{fig:angles}.
One has
\begin{equation}
  \label{eq:phase_ij}
   \cot \phi_{ij}=\frac{\cos\theta_i\sin\theta_j-\sin\theta_i\cos\theta_j\cos(\phi_i-\phi_j)}
{\sin\theta_i\sin(\phi_i-\phi_j)},
\end{equation}
or
\begin{equation}
  \label{eq:phase_ij_in_z}
  \phi_{ij}(z_i,z_j)=\frac{1}{2i}\ln\frac{(\bar z_i z_j+1)(z_i-z_j)\bar z_j}{(z_i\bar z_j+1)(\bar z_i-\bar z_j)z_j}.
\end{equation}

This function is harmonic with respect to $z_i$ and $z_j$. The Cauchy Riemann
equations yield the ``short distance'' behavior
\begin{equation}
  \label{}
  |F_{ij}|=\sqrt{\frac{(1+z_i\bar z_j)(1+\bar z_i z_j)}{ (z_i-z_j)(\bar z_i-\bar z_j)}},
\end{equation}
leading to the function
\begin{equation}
  \label{eq:F_ij}
  F_{ij}=\sqrt{z_j\bar z_j}\frac{\bar z_i+\frac{1}{z_j}}{\bar z_i-\bar z_j},
\end{equation}
which, in the  limiting case $(z_i=z, z_j\rightarrow 0)$,  coincides
with $F=1/\bar z$ (up to a constant shift in the phase).
Clearly, this function describes    a vortex-antivortex pair
on the
projected plane, at position $z_j$ and $-1/\bar z_j$, which are, not surprisingly, the
positions where the vortex pierces  the sphere.
The Aharonov-Bohm Hamiltonian follows
from the free one $H^o_i$ for particle $i$, which is given by
\begin{equation}
  \label{eq:H0_stereographic}
  H^o_i=-\frac{1}{2}(1+z_i\bar z_i)^2 \partial_i \bar \partial_i,
\end{equation}
and the substitution of  the phase $\exp(\phi_{ij}(z_i,z_j))$ in $\Psi^o_i$.
Rather, one can
directly construct the Hamiltonian $\tilde H_i$
by substitution of the function (\ref{eq:F_ij}), i.e.,
$\Psi^o_i=(F_{ij})^\alpha\tilde\Psi_i$

\begin{align}
  \label{eq:H_tilde}
  \tilde H_i &= F_{ij}^{-\alpha} H^o_i F_{ij}^\alpha \nonumber\\
&=-\frac{1}{2}(1+z_i\bar z_i)^2\Bigl\{\partial_i\bar\partial_i +\alpha (\partial_i \ln F_i) \bar \partial_i
+ \alpha (\bar \partial_i \ln F_i) \partial_i + \alpha^2(\partial_i F_i)(\bar\partial_i F_i)
+ \alpha \partial_i \bar\partial_i F_i\Bigr\}\nonumber\\
&=-\frac{1}{2}(1+z_i\bar z_i)^2\left\{\partial_i\bar\partial_i - \alpha \frac{1+z_j \bar z_j}{(1+\bar z_i z_j)(\bar z_i -\bar z_j)}\partial_i\right\}.
\end{align}
The eigenfunctions are deduced from (\ref{ouf},\ref{oufbis}) replacing $F$
by $F_{ij}$ as well as the argument of the hypergeometric function $x$ by
$x_{ij}$ given by
 \begin{equation}
  \label{eq:cos_theta_ij}
  x_{ij}=\cos\theta_{ij}=\frac{2|z_i-z_j|^2}{(1+z_i\bar z_i)(1+z_j\bar z_j)} - 1.
\end{equation}
We note that

\begin{equation}
  \label{eq:relations_x} \frac{1+x_{ij}}{2}=\frac{|z_i-z_j|^2}{(1+z_i\bar
  z_i)(1+z_j\bar z_j)},\;\text{ and }\;\frac{1-x_{ij}}{2}=\frac{|1+z_i\bar
  z_j|^2}{(1+z_i\bar z_i)(1+z_j\bar z_j)}.
\end{equation}
A particular subset of the solutions possesses again a simple analytic form
when  $\lambda=\mu$, i.e., $k=0$

\begin{align}
  \label{eq:one_body_particular} \Psi^o_i = \left\{
\begin{array}{lll}
(F_{ij})^{m+\alpha} \left( \frac{1+x_{ij}}{2}\right)^{m+\alpha}&= \left(\frac{(1+
\bar z_i z_j)(z_i-z_j)}{(1+z_i\bar z_i)(1+z_j\bar z_j)}\right)^{m+\alpha},&
m+\alpha>0\\ (\bar F_{ij})^{-m-\alpha} \left( \frac{1+x_{ij}}{2}\right)^{-m-\alpha}&=
\left(\frac{(1+ z_i \bar z_j)(\bar z_i-\bar z_j)}{(1+z_i\bar z_i)(1+z_j\bar
z_j)}\right)^{-m-\alpha},& m+\alpha<0,\\
\end{array}
\right.
\end{align}
which are deformations of (\ref{u}) and (\ref{baru}). As previously, and
taking advantage of (\ref{eq:one_body_particular}), one can define a Hamiltonian 
$\tilde H_{i}'$ with the prescription $\Psi^o_i=\left(\frac{(1+ \bar z_i
z_j)(z_i-z_j)}{(1+z_i\bar z_i)(1+z_j\bar z_j)}\right)^{\alpha}\tilde\Psi'_i$
giving
\begin{equation}
  \label{eq:Htildeprimei} \tilde H'_i = -\frac{1}{2}(1+z_i \bar z_i)^2
  \left\{\partial_i \bar\partial_i+\alpha\frac{1+\bar z_i z_j}{(1+z_i \bar
  z_i)(z_i-z_j)}\bar \partial_i-\alpha\frac{z_i-z_j}{(1+z_i \bar z_i)(1+\bar
  z_i z_j)}\partial_i\right\}+\frac{1}{2}\alpha(\alpha+1).
\end{equation}

\subsection{Generalization to the N-body case}
\label{sec:N_body}

Consider now a system of $N$ identical particles of charge $e$ and attached
flux tubes $\alpha\phi_0$ (now with $\alpha\in[-1,1]$) piercing the sphere at
the positions of the particles and exiting at their antipodes. In this
spherical model, the total flux through the sphere is null, thus
there is no Dirac quantization condition on $\alpha$. Locally, the 
relative phase of two particles (\ref{eq:phase_ij}) is anyon-like, however globally --
and in contrast to the planar anyon model, where $\phi_{ij}=\phi_{ji}+\pi$ -- here the phase
between two particles is not symmetric. Consequently, it is not possible to
write a global phase for the many-particle wave function. Nonetheless, 
each particle must see all the fluxes carried by the other particles,
thus the phase for particle $i$ is
\begin{equation}
  \label{eq:general_phi}
  \phi_i= \sum_{j\neq i} \phi_{ij}.
\end{equation}
Consequently, the A-B Hamiltonian $H_i^{\mathrm AB}$ for particle $i$ coupled to all other
particles follows as usual from $H^o_i$ by extracting the multivalued phase
$\exp(i\alpha\phi_i)$. The global A-B Hamiltonian of the system is obtained as
the sum of the $H_i^{\mathrm AB}$'s
\begin{equation}
  \label{eq:H_N_body_AB}
  H^{\mathrm AB} = \sum_i H_i^{\mathrm AB}.
\end{equation}
Contrary to the phase, the ``short-distance'' behavior, i.e., the
absolute value of (\ref{eq:F_ij}), is symmetric under the exchange of
particles, $|F_{ij}|=|F_{ji}|$, as it should for a ``distance''. One can thus
substitute in the A-B Hamiltonian $H^{\mathrm AB}$ a global ``short-distance'' behavior
$\prod_{i<j}|F_{ij}|^{\alpha}$ (here one restricts to $\alpha\in[0,1]$) to obtain a global $\tilde H$ Hamiltonian.
Equivalently, on can directly start from the free Hamiltonian $H^o_i$
for particle $i$ and substitute $\prod_{j, j\ne i}F_{ij}^{\alpha}$ to obtain
\begin{equation}
\label{eq:H_tilde_N}
\tilde H_i=-\frac{1}{2}(1+z_i \bar z_i)^2\left\{\partial_i
\bar\partial_i-\alpha\sum_{j\neq i}\frac{1+z_j\bar z_j}{(1+\bar z_iz_j)(\bar
z_i - \bar z_j)}\partial_i\right\}.
\end{equation}
Then $\tilde H=\sum_i \tilde H_i$.
Or, following the lines which lead to (\ref{eq:Htildeprimei}), substitute
$\prod_{j, j\ne i}\left(\frac{(1+
\bar z_i z_j)(z_i-z_j)}{(1+z_i\bar z_i)(1+z_j\bar z_j)}\right)^{\alpha}$
to obtain $\tilde H'_i$, then
\begin{align}
  \label{eq:H_tilde_N_prime}
  &\tilde H'_N=\sum_i \tilde H'_i=\nonumber\\
&-\frac{1}{2}\sum_i^N (1+z_i\bar z_i)^2\left\{
      \partial_i \bar\partial_i
    + \alpha \sum_{j\neq i} \frac{1+\bar z_i z_j}{(z_i-z_j)(1+z_i\bar z_i)}\bar \partial_i
    - \alpha \sum_{j\neq i} \frac{z_i-z_j}{(1+\bar z_i z_j)(1+z_i\bar z_i)} \partial_i \right\}
 \nonumber\\
    & +\frac{\alpha(\alpha+1)}{2} N(N-1) + \frac{1}{2}\alpha^2
\sum_{i}\sum_{\substack{j,k \\j\neq i,\,k\notin \{i,j\}}}\frac{( 1+\bar z_i z_k)(z_i-z_j)}
{(z_i-z_k)(1+\bar z_i z_j)}.
\end{align}
This Hamiltonian has a complicated structure, as can be seen in particular 
for the 3-body $\alpha^2$ term.

\subsection{The two-body case and its ground state}
In the 2-anyon case things simplify, since the $\alpha^2$ term is now merely a c-number:
\label{sec:two_body}
\begin{align}
  \label{eq:H_tilde_N2}
  \tilde H'_2 &= -\frac{1}{2}(1+z_1 \bar z_1)^2 \left\{\partial_1 \bar\partial_1+\alpha\frac{1+\bar z_1 z_2}{(1+z_1 \bar z_1)(z_1-z_2)}\bar \partial_1-\alpha\frac{z_1-z_2}{(1+z_1 \bar z_1)(1+\bar z_1 z_2)}\partial_1\right\}\nonumber\\
&-\frac{1}{2}(1+z_2 \bar z_2)^2 \left\{\partial_2 \bar\partial_2+\alpha \frac{1+\bar z_2 z_1}{(1+z_2 \bar z_2)(z_2-z_1)}\bar \partial_2- \alpha\frac{z_2-z_1}{(1+z_2 \bar z_2)(1+\bar z_2 z_1)}\partial_2\right\}+\alpha(\alpha+1)
\end{align}
It is immediate
that the ground state is $\tilde\Psi_{GS}=1$, with energy $E_{GS}=\alpha(\alpha+1)$,
i.e., for  the A-B Hamiltonian
\begin{equation}
  \label{eq:GS_AB_N2}
  \Psi_{GS}^{\mathrm AB}=\left(\frac{|1+ \bar z_1 z_2| |z_1-z_2|}
{(1+z_1\bar z_1)(1+z_2\bar z_2)}\right)^\alpha.
\end{equation}
Note that when $\alpha\to 1$, the 2-body wave function has a fermionic behavior,
since, when $z_1\simeq z_2$, it vanishes as $|z_1-z_2|$ (also when $z_1\simeq
-1/\bar z_2$ it vanishes as $|1+ \bar z_1 z_2|$). Likewise, note that 
$\phi_{12}\simeq \phi_{21}$ up to a constant, which confirms that the model is locally
anyon-like.

Apart from the ground state, no other particular solutions are known so far.
Among other approaches,
\footnote{e.g., another possible approach to the search of eigenstates of the problem 
is a different rewriting of  $\tilde H'_2$ in analogy to the anyon problem on the plane, where
one may separate an angular momentum part from an interacting 2-body term :
$$
  \label{eq:alternative_2-body}
  \tilde H'_2 = \protect\underbrace{-\frac{1}{2}(1+z_1 \bar z_1)^2 \left\{
\partial_1 \bar\partial_1 -\alpha \frac{z_1 \partial_1 + \bar z_1 \bar \partial_1}
{1+z_1 \bar z_1}\right.}_{\text{one-body, pseudo-angular-momentum
part}} \left. + \alpha \frac{1}{z_1-z_2}\bar \partial_1 + \alpha
\frac{1}{\bar z_1 + \frac{1}{z_2}}\partial_1\right\} +
(1\leftrightarrow 2) + \alpha(\alpha+1)
$$
} we explored whether there are eigenstates
based on the set of solutions of the one-body problem.

\section{Conclusion}
\label{sec:conclusion}
A dimensional reduction scheme was proposed, which relates a quantum
mechanical free particle on the sphere of radius $R$ to a chiral particle on a
circle of same radius. This projection parallels a somewhat analogous
projection between a 2d particle in a harmonic well and a 1d particle in a
harmonic well. Further a generalized Aharonov-Bohm model on the sphere was defined, with in mind its
dimensional reduction to the Calogero-Sutherland model. This model 
was found to have an anyon-like character and has
interesting properties in that it allows for continuous
values of its coupling parameter $\alpha$, since it is not subject to a Dirac quantization condition.
Furthermore, the corresponding $N$-body Hamiltonian may be expressed in a 
simple form in the stereographic projection coordinates. Its construction follows a geometrical
analogy with the one on the plane. However, and contrary to the plane where
the anyon model is defined from a free Hamiltonian by a global multi-valued
statistical phase, here the phase of each particle accounts for its
interactions with the flux lines attached to the other particles. Adding up the
vector potentials in the interacting description yields the Aharonov-Bohm Hamiltonian.
This model does not share all the usual properties of anyon exchange
statistics due to a global phase asymmetry $\phi_{ij}\neq\phi_{ji}$; in particular the limit 
$\alpha=1$ does not correspond to free fermions. 
Yet, at short distance, the phase is symmetric $\phi_{ij}\simeq \phi_{ji}$ and this locally anyon-like
character is also reflected in the fermionic short distance behavior for the 2-body ground state.

Despite its relative simplicity, eigenstates of the model seem rather complicated to construct in the general case. 
Consequently, the question whether the proposed model allows for a 
dimensional reduction towards the Calogero-Sutherland model is still open.

\section*{Acknowledgements}
\label{sec:acknowlege}
We thank Alain Comtet for suggesting  the question 
of the relation between the anyon and  Calogero-Sutherland models, 
as well as for his
participation in the early stages of this work. Likewise, we 
thank  Jean Desbois for helpful discussions.
S. M. acknowledges the hospitality of the Laboratoire de Physique 
Th\'eorique et Mod\`eles Statistiques Orsay and support by
RFBR grant N 04-02-17087.


\end{document}